\documentstyle[prd,aps,preprint,tighten,epsfig]{revtex}
\begin{document}
\draft

\title{Possible Capture of keV Sterile Neutrino Dark Matter \\
on Radioactive $\beta$-decaying Nuclei}
\author{{\bf Y.F. Li} ~ and ~ {\bf Zhi-zhong Xing}}
\address{Institute of High Energy Physics, Chinese Academy of Science,
Beijing 100049, China \\
({\it Electronic address: liyufeng@ihep.ac.cn, xingzz@ihep.ac.cn})}
\maketitle

\begin{abstract}
There exists an observed ``desert" spanning six orders of magnitude
between ${\cal O}(0.5)$ eV and ${\cal O}(0.5)$ MeV in the fermion
mass spectrum. We argue that it might accommodate one or more keV
sterile neutrinos as a natural candidate for warm dark matter. To
illustrate this point of view, we simply assume that there is one keV
sterile neutrino $\nu^{}_4$ and its flavor eigenstate $\nu^{}_s$ weakly
mixes with three active neutrinos. We clarify different active-sterile
neutrino mixing factors for the radiative decay of $\nu^{}_4$ and
$\beta$ decays in a self-consistent
parametrization. A direct detection of this keV sterile
neutrino dark matter in the laboratory is in principle possible since
the $\nu^{}_4$ component of $\nu^{}_e$ can leave a distinct imprint on
the electron energy spectrum when it is captured on radioactive
$\beta$-decaying nuclei. We carry out an analysis of its signatures in
the capture reactions $\nu^{}_e + ~^3{\rm H} \to ~^3{\rm He} + e^-$ and
$\nu^{}_e + ~^{106}{\rm Ru} \to ~^{106}{\rm Rh} + e^-$ against the
$\beta$-decay backgrounds, and conclude that this experimental approach
might not be hopeless in the long run.
\end{abstract}


\newpage

\framebox{\bf 1} ~
The existence of dark matter (DM) in the Universe has been established,
but its nature remains a fundamental puzzle in particle physics and
cosmology. Within the standard model (SM) three active neutrinos and
their antiparticles may constitute hot DM, which only has a tiny
contribution to the total matter density of the Universe. A careful study
of the structure formation indicates that most DM should be cold
at the onset of the galaxy formation
\cite{PDG}. Possible candidates for cold DM include weakly interacting
massive particles, axions and other exotic objects beyond the SM \cite{Feng}.
Between the hot and cold limits, warm DM is another possibility of
accounting for the observed non-luminous and non-baryonic matter content
in the Universe. Its presence may solve or soften several
problems that one has so far encountered in the DM simulations \cite{Bode}
(e.g., damping the inhomogeneities on small scales by reducing the number 
of dwarf galaxies or smoothing the cusps in the DM halos). 
Sterile neutrinos are expected to be a good
candidate for warm DM, if their masses are of ${\cal O}(1)$ keV and
their lifetimes are much longer than the age of the Universe \cite{Review}.
They could be produced in the early Universe in several
ways (e.g., either via non-resonant active-sterile neutrino oscillations \cite{DW} or via resonant active-sterile neutrino oscillations in the
presence of a non-negligible lepton number asymmetry \cite{Shi}). DM in the
form of keV sterile neutrinos may not only suppress the formation of dwarf
galaxies and other small-scale structures but also have impacts on the
X-ray spectrum, the velocity distribution of pulsars and the formation
of the first stars \cite{Kusenko1}. Hence their masses and mixing angles can
get stringent constraints from the measurements of the X-ray fluxes and
the Lyman-$\alpha$ forest \cite{Abazajian}.

Sterile neutrinos of ${\cal O}(1)$ keV are well motivated in some
theoretical models. Two typical examples of this kind are
the $\nu$MSM \cite{Shaposhnikov} and the {\it split seesaw} model
\cite{Yanagida}, which can realize the
seesaw and leptogenesis ideas and accommodate one keV sterile
neutrino as the DM candidate. Other interesting scenarios
have also been proposed \cite{Kusenko2}. Here we give a purely
{\it phenomenological} and {\it model-independent} argument to support
the conjecture of keV sterile neutrinos as warm DM. Note that there
exists an apparent ``desert" spanning six orders of magnitude
between ${\cal O}(0.5)$ eV and ${\cal O}(0.5)$ MeV in the SM fermion
mass spectrum as shown in FIG. 1. Such a puzzle is a part of the
flavor problem in the SM, and it might be solved if there exist
one or more keV sterile neutrinos in the desert as a natural candidate
for warm DM \cite{XING}
\footnote{One of us (Z.Z.X.) first put forward this simple point of view as
a positive comment on Alexander Kusenko's talk entitled ``The dark side
of the light fermions" at the 16th Yukawa International
Seminar (YKIS) Symposium on Particle Physics beyond the Standard Model,
Kyoto, March 2009.}.
For simplicity, we assume that only a single sterile neutrino $\nu^{}_s$
hides in the desert and it weakly mixes with three active neutrinos.
We also assume that its mass eigenstate $\nu^{}_4$ possesses a rest
mass of ${\cal O}(1)$ keV and satisfies all the prerequisites of warm DM.
We shall focus on how to directly detect this sterile
neutrino DM in the laboratory.

Because $\nu^{}_s$ mixes with $\nu^{}_e$, a careful study of the kinematics
of different $\beta$ decays is in principle possible to probe keV
sterile neutrino DM \cite{Shaposhnikov2}. A more promising method, which is
quite analogous to the direct detection of active
\cite{Weinberg,Cocco,Blennow,Li} and sterile
\cite{Li} components of the cosmic neutrino background (C$\nu$B), is to
investigate the capture of $\nu^{}_4$ on radioactive $\beta$-decaying
nuclei \cite{Liao}. The point is simply that the $\nu^{}_4$ component
of $\nu^{}_e$ can leave a distinct imprint on
the electron energy spectrum when it is captured on a nucleus.
We carry out an analysis of the signatures of this keV sterile neutrino DM
in the capture reactions $\nu^{}_e + ~^3{\rm H}
\to ~^3{\rm He} + e^-$ and $\nu^{}_e + ~^{106}{\rm Ru} \to ~^{106}{\rm Rh}
+ e^-$ against the $\beta$-decay backgrounds to illustrate the
salient features of such a detection method. Our work is
different from Ref. \cite{Liao} in several aspects: (1) we are not
subject to any specific neutrino mass models such as the $\nu$MSM; (2)
we clarify different active-sterile neutrino mixing factors for the radiative
decay of $\nu^{}_4$ and $\beta$ decays, which correspond to the
constraints from measurements of the X-ray spectrum and the $\nu^{}_4$
capture, in a self-consistent parametrization; (3) we take account of
the finite energy resolution in detecting the electron energy
spectrum to make this method more realistic; and (4) we take into account
the finite lifetimes of $^3{\rm H}$ and $^{106}{\rm Ru}$ and find that the
latter may have a non-negligible effect on the capture rate of $\nu^{}_4$
on $^{106}{\rm Ru}$ nuclei.

\vspace{0.1cm}

\framebox{\bf 2} ~
For simplicity, we consider a SM-like electroweak theory in which
there exists one sterile neutrino ($\nu^{}_s$) belonging to the
isosinglets together with three charged leptons ($e$, $\mu$, $\tau$)
and three active neutrinos
($\nu^{}_e$, $\nu^{}_\mu$, $\nu^{}_\tau$) belonging to the isodoublets.
Without loss of generality, one may choose to identify the flavor
eigenstates of charged leptons with their mass eigenstates. In this
flavor basis the $3 \times 4$ lepton mixing matrix appearing in
the Lagrangian of weak charged-current interactions
links the neutrino flavor eigenstates $\nu^{}_\alpha$
(for $\alpha = e, \mu, \tau$) to the neutrino mass eigenstates $\nu^{}_i$
(for $i = 1, 2, 3, 4$). Although $\nu^{}_s$ does not directly
participate in the standard weak interactions, it
may oscillate with active neutrinos or interact with matter in an
indirect way (i.e., via its mixing with $\nu^{}_\alpha$).
So we are concerned about the $4 \times 4$ active-sterile
neutrino mixing matrix $V$ whose explicit form reads
\begin{equation}
\left ( \matrix{ \nu^{}_e \cr \nu^{}_\mu \cr
\nu^{}_\tau \cr \nu^{}_s \cr} \right ) \; = \;
\left ( \matrix{ V^{}_{e1} & V^{}_{e2} & V^{}_{e3} & V^{}_{e4} \cr
V^{}_{\mu 1} & V^{}_{\mu 2} & V^{}_{\mu 3} & V^{}_{\mu 4} \cr
V^{}_{\tau 1} & V^{}_{\tau 2} & V^{}_{\tau 3} & V^{}_{\tau 4}
\cr V^{}_{s 1} & V^{}_{s 2} & V^{}_{s 3} & V^{}_{s 4} \cr} \right )
\left ( \matrix{ \nu^{}_1 \cr \nu^{}_2 \cr \nu^{}_3 \cr \nu^{}_4
\cr} \right ) \; ,
\end{equation}
where $\nu^{}_4$ denotes the mass eigenstate of $\nu^{}_s$. A
generic parametrization of $V$ needs six mixing angles $\theta^{}_{ij}$
and three (Dirac) or six (Majorana) CP-violating phases
$\delta^{}_{ij}$ (for $1 \leq i < j \leq 4$).
In the assumption of $|\theta^{}_{i4}| \ll 1$ (for $i=1,2,3$), one
may simplify the standard parametrization of $V$ advocated in
Ref. \cite{Guo}. For instance, $V^{}_{s1} \simeq
\hat{s}^*_{14}$, $V^{}_{s2} \simeq \hat{s}^*_{24}$, $V^{}_{s3}
\simeq \hat{s}^*_{34}$ and $V^{}_{s4} \simeq 1$ for the matrix
elements relevant to $\nu^{}_s$, where
$\hat{s}^{}_{ij} \equiv s^{}_{ij} \ e^{{\rm i}\delta^{}_{ij}}$
and $s^{}_{ij} \equiv \sin \theta^{}_{ij}$ are defined.
The dominant decay mode of $\nu^{}_4$ is
$\nu^{}_4 \to \nu^{}_\alpha + \nu^{}_\beta + \overline{\nu}^{}_\beta$
(for $\alpha, \beta = e, \mu, \tau$) mediated by the $Z^0$ boson at the
tree level, and its rate is given by
\begin{equation}
\sum^\tau_{\alpha =e} \sum^\tau_{\beta =e}
\Gamma (\nu^{}_4 \to \nu^{}_\alpha + \nu^{}_\beta +
\overline{\nu}^{}_\beta) \; =\; \frac{C^{}_\nu G^2_{\rm F}
m^5_4}{192 \pi^3} \sum^\tau_{\alpha = e} |V^{}_{\alpha 4}|^2
\; =\; \frac{C^{}_\nu G^2_{\rm F} m^5_4}{192 \pi^3}
\sum^3_{i=1} |V^{}_{s i}|^2 \; ,
\end{equation}
where $C^{}_\nu =1$ for the Dirac neutrinos or $C^{}_\nu = 2$ for
the Majorana neutrinos, $m^{}_4$ denotes the mass of $\nu^{}_4$,
and $m^{}_4 \gg m^{}_i$ (for $i=1,2,3$) holds.
The lifetime of $\nu^{}_4$ turns out to be
\begin{equation}
\tau^{}_{\nu^{}_4} \; \simeq \; \frac{2.88 \times 10^{27}}{C^{}_\nu}
\left(\frac{m^{}_4} {1 \ {\rm keV}} \right)^{-5} \left(\frac{\displaystyle
s^2_{14}+s^2_{24}+s^2_{34}}{10^{-8}}\right)^{-1} {\rm s} \; ,
\end{equation}
which can be much larger than the age of the Universe ($\sim
10^{17}$ s). So keV sterile neutrinos may be a natural candidate
for warm DM.

We are more interested in the subdominant decay channel of
$\nu^{}_4$ --- its radiative decay at the one-loop level.
Following Ref. \cite{Pal} and taking $m^{}_4 \gg
m^{}_i$ (for $i=1,2,3$), we find
\begin{eqnarray}
\sum^3_{i=1} \Gamma (\nu^{}_4 \to \nu^{}_i + \gamma) & \simeq &
\frac{9 \alpha^{}_{\rm em} C^{}_\nu G^2_{\rm F} m^5_4}{512 \pi^4}
\sum^3_{i=1} \left| \sum^\tau_{\alpha =e} V^{}_{\alpha 4} V^*_{\alpha i}
\right|^2 \; = \; \frac{9 \alpha^{}_{\rm em} C^{}_\nu G^2_{\rm F} m^5_4}{512
\pi^4} \sum^3_{i=1} \left| V^{}_{s 4} V^*_{s i} \right|^2 \nonumber
\\
& \simeq & \frac{9 \alpha^{}_{\rm em} C^{}_\nu G^2_{\rm F} m^5_4}{512 \pi^4}
\left (s^2_{14}+s^2_{24}+s^2_{34} \right) \; ,
\end{eqnarray}
where $\alpha^{}_{\rm em} \simeq 1/137$ denotes the electromagnetic
fine-structure constant, and $C^{}_\nu = 1$ (Dirac) or 2 (Majorana).
A comparison of Eq. (4) with Eq. (2) yields $27 \alpha^{}_{\rm em}
/(8\pi) \simeq 1/128$ for the ratio of two decay rates \cite{Barger}.
A search for the X-ray flux arising from the radiative decay of $\nu^{}_4$
can set a model-independent bound on its mass and mixing angles \cite{Review}:
\begin{equation}
s^2_{14}+s^2_{24}+s^2_{34} \; \lesssim \;
1.8 \times 10^{-5} \left(\frac{1 \ {\rm keV}}{m^{}_4}\right)^5 \; .
\end{equation}
Another model-independent constraint on $m^{}_4$ comes from the
Tremaine-Gunn bound on DM particles \cite{TGbound}:
$m^{}_4 \gtrsim 0.4 ~{\rm keV}$. More stringent
constraints on the mass and mixing parameters of keV sterile
neutrinos can be achieved when their production mechanism and
proper DM abundance are taken into account.
In the non-resonant active-sterile neutrino oscillation
scenario \cite{DW}, for instance, one obtains
$1.7 ~{\rm keV} < m^{}_4 < 6.3 ~{\rm keV}$ \cite{Abazajian,TG22}
\footnote{Here the upper bound on $m^{}_4$ comes from the X-ray
measurement \cite{Abazajian}, which is in contradiction with
$m^{}_4 > 8.0$ keV obtained from the Lyman-$\alpha$
forest \cite{lyman}. But in the entropy dilution scenario the
Lyman-$\alpha$ bound on $m^{}_4$ can even be lowered to
$1.6$ keV \cite{Liao,lindner}.}
together with \cite{Review}
\begin{equation}
s^2_{14}+s^2_{24}+s^2_{34} \; \simeq \; 3.2 \times 10^{-9}
\left(\frac{3 \ {\rm keV}}{m^{}_4}\right)^{1.8} \;
\end{equation}
to fit the observed DM abundance (i.e., $\Omega^{}_{\rm DM}
\simeq 0.21$ \cite{PDG}) in the Universe at the present time.

Note that an analysis of recent observational data on the
Willman 1 dwarf spheroidal galaxy in the Chandra X-ray Observatory
shows evidence of a relic sterile neutrino with $m^{}_4 \sim
5$ keV and $s^2_{14} + s^2_{24} + s^2_{34} \sim 10^{-9}$
\cite{Kusenko-DM}. On the other hand, the excess of the intensity in the
FeXXVI Lyman-$\gamma$ line in the spectrum of the Galactic center
observed recently by the Suzaku X-ray mission hints at the existence of
a relic sterile neutrino with $m^{}_4 \sim 17.4$ keV and
$s^2_{14} + s^2_{24} + s^2_{34} \sim 10^{-12}$ \cite{Silk}. In
either case the inferred number density of sterile neutrinos can
account for a part or all of DM.  Although the statistical significance
of the above evidence for keV sterile neutrino DM is not sufficiently
strong, it {\it does} motivate us to take the issue more seriously
than before.

\vspace{0.1cm}

\framebox{\bf 3} ~
Now we look at how to directly detect keV sterile neutrino DM in the
laboratory. We focus our attention on the capture of cosmic
low-energy electron neutrinos on radioactive $\beta$-decaying nuclei
(i.e., $\nu^{}_e + N \to N^\prime + e^-$), in which the $\nu^{}_4$
component of $\nu^{}_e$ may leave a distinct imprint on the electron
energy spectrum. This capture reaction can happen for any kinetic
energy of the incident neutrino, because the corresponding $\beta$
decay $N \to N^\prime + e^- + \overline{\nu}^{}_e$ always
releases some energies ($Q^{}_\beta = m^{}_N - m^{}_{N^\prime} - m^{}_e
> 0$). So it has a unique advantage in detecting cosmic neutrinos
with $m^{}_i \ll Q^{}_\beta$ and extremely low energies
\cite{Weinberg,Cocco,Blennow,Li}. In the low-energy limit the
product of the cross section of non-relativistic neutrinos
$\sigma^{}_{\nu^{}_i}$ and the neutrino velocity $v^{}_{\nu^{}_i}$
converges to a constant value \cite{Cocco}, and thus the capture
rate for each $\nu^{}_{i}$ reads
\begin{equation}
{\cal N}^{}_{\nu^{}_i} \; = N^{}_{\rm T} |V^{}_{ei}|^2 \sigma^{}_{\nu^{}_i}
v^{}_{\nu^{}_i} n^{}_{\nu^{}_i} \; ,
\end{equation}
where $n^{}_{\nu^{}_i}$ denotes the number density of $\nu^{}_{i}$
around the Earth or in our solar system, and $N^{}_{\rm T}$ measures
the average number of target nuclei for the duration of detection.
The standard Big Bang model predicts $n^{}_{\nu^{}_i} =
n^{}_{\overline{\nu}^{}_i} \simeq 56 ~{\rm cm}^{-3}$ (for $i=1,2,3$)
today for each species of active neutrinos. As for the keV sterile
neutrinos, we assume that they were produced in the early Universe
through active-sterile neutrino oscillations \cite{DW,Shi} and their
number density could account for the total amount of DM. With the
help of the average density of DM in our Galactic neighborhood
(i.e., $\rho^{\rm local}_{\rm DM} \simeq 0.3 ~{\rm GeV} \ {\rm
cm}^{-3}$ \cite{Kamionkowski}), one may estimate the number density
of $\nu^{}_4$ to be $n^{}_{\nu^{}_4} \simeq 10^{5} \ (3 ~{\rm
keV}/m^{}_4) ~{\rm cm}^{-3}$. The average number of target nuclei in
the detecting time interval $t$ can be calculated as follows:
\begin{equation}
N^{}_{\rm T} \; =\; \frac{1}{t}\int_0^{t}N(0) \ e^{-\lambda
t^\prime}{\rm d}t^\prime \; =\; \frac{N(0)}{\lambda\,t} \left(1 -
e^{-\lambda t} \right) \; ,
\end{equation}
where $\lambda={\rm ln}2/t^{}_{1/2}$ with $t^{}_{1/2}$ being the
half-life of each target nucleus, and $N(0)$ is the initial number
of target nuclei. In the capture reaction each non-relativistic
$\nu^{}_i$ can in principle produce a monoenergetic electron with
the kinetic energy
$T^{(i)}_{e} = Q^{}_\beta + E^{}_{\nu^{}_i} \simeq Q^{}_\beta + m^{}_{i}$.
Because a realistic experiment must be subject to a finite energy
resolution, we consider the Gaussian energy resolution function
\begin{equation}
R\left(T^{}_{e}, T^{(i)}_{e}\right) \; = \; \frac{1}{\sqrt{2\pi} \,\sigma}
\exp\left[-\frac{\left(T^{}_{e} - T^{(i)}_{e}\right)^2}{2\sigma^2}
\right] \; ,
\end{equation}
which makes the ideally discrete energy lines of the electrons
to spread and form a continuous energy spectrum.
Then the overall neutrino capture rate (i.e., the energy spectrum
of the detected electrons) is given by
\begin{equation}
{\cal N}^{}_\nu \; = \; \sum^4_{i=1} {\cal N}^{}_{\nu^{}_i} \,
R\left(T^{}_{e}, T^{(i)}_{e}\right) \; = \; \sum^4_{i=1} N^{}_{\rm T}
|V^{}_{ei}|^2 \sigma^{}_{\nu^{}_i} v^{}_{\nu^{}_i} n^{}_{\nu^{}_i} \,
R\left(T^{}_{e}, T^{(i)}_{e}\right) \; .
\end{equation}
It is worth emphasizing that $N^{}_{\rm T} \simeq N(0)$ is an
excellent approximation for $^3{\rm H}$ \cite{Li},
but it will not be true for $^{106}{\rm Ru}$ and some other heavy nuclei
with $t^{}_{1/2} \sim t$ or $t^{}_{1/2} < t$.

The main background of a neutrino capture process is its
corresponding $\beta$ decay. The finite energy resolution may
push the outgoing electron's ideal endpoint $Q^{}_\beta - {\rm
min}(m^{}_i)$ towards a higher energy region, and hence it is
possible to mimic the desired signal of the neutrino capture
reaction. Given the same energy resolution as that in Eq. (9),
we can describe the energy spectrum of a $\beta$ decay as
\begin{eqnarray}
\frac{{\rm d} {\cal N}^{}_\beta}{{\rm d}T^{}_e} & = &
\int_0^{Q^{}_{\beta}- {\rm min}(m^{}_i)} {\rm d} T^\prime_e \,
\left\{ N^{}_{\rm T} \, \frac{G^2_{\rm F} \, \cos^2\theta^{}_{\rm
C}}{2\pi^3} \, F\left(Z, E^{}_{e}\right) \, |{\cal M}|^2 \sqrt{E^2_e -
m^2_e} \, E^{}_{e} \left(Q^{}_{\beta} - T^\prime_e\right) \right .
\nonumber \\
& & \left . \times \sum^4_{i=1} \left[ |V^{}_{ei}|^2\sqrt{\left(Q^{}_{\beta}-
T^\prime_e \right)^2 - m_i^2} ~ \Theta\left(Q^{}_{\beta} - T^\prime_e -
m^{}_i\right) \right] R\left(T^{}_e, T^\prime_e\right) \right\} \; ,
\end{eqnarray}
where $T^\prime_e = E^{}_e - m^{}_e$ is the intrinsic kinetic
energy of the outgoing electron, $F(Z, E^{}_{e})$ denotes the Fermi
function, $|{\cal M}|^2$ stands for the dimensionless contribution
of relevant nuclear matrix elements \cite{Weinheimer}, and
$\theta^{}_{\rm C} \simeq 13^\circ$ is the Cabibbo angle.

Let us stress that the active-sterile neutrino mixing factor
in Eq. (10) or Eq. (11) is very different from that in the radiative
decay of $\nu^{}_4$ as given by Eq. (4). Taking account of
$|\theta^{}_{i4}| \ll 1$ (for $i=1,2,3$), we have
$|V^{}_{e1}| \simeq c^{}_{12} c^{}_{13}$, $|V^{}_{e2}| \simeq
s^{}_{12} c^{}_{13}$ and $|V^{}_{e3}| \simeq s^{}_{13}$ to a good
degree of accuracy in the standard parametrization of $V$ \cite{Guo}.
Nevertheless,
\begin{eqnarray}
|V^{}_{e4}|^2 & \simeq & \left| c^{}_{12} c^{}_{13} \hat{s}_{14}
+ \hat{s}^*_{12} c^{}_{13} \hat{s}^{}_{24} + \hat{s}^*_{13}
\hat{s}^{}_{34} \right|^2 \nonumber \\
& \simeq & c^2_{12} s^2_{14} + s^2_{12} s^2_{24} +
2 c^{}_{12} s^{}_{12} s^{}_{14} s^{}_{24} \cos \left( \delta^{}_{24}
-\delta^{}_{12} -\delta^{}_{14} \right)
\end{eqnarray}
holds for $s^{}_{13} \ll 1$ and $c^{}_{13} \simeq 1$. We see that
the neutrino mixing parameters appearing in $\nu^{}_e + N \to
N^\prime + e^-$ and $N \to N^\prime + e^- + \overline{\nu}^{}_e$ are
mainly $\theta^{}_{12}$, $\theta^{}_{14}$, $\theta^{}_{24}$ and
$\delta^{}_{24} -\delta^{}_{12}-\delta^{}_{14}$. While
$\theta^{}_{12}$ is already known from the solar neutrino
oscillation experiments \cite{PDG}, the others are relevant to the
sterile neutrino $\nu^{}_4$ and thus undetermined. In particular,
the CP-violating phase $\delta^{}_{24}
-\delta^{}_{12}-\delta^{}_{14}$ contributes to the neutrino capture
reactions and the corresponding $\beta$ decays. In comparison, the
rate of the radiative decay of $\nu^{}_4$ given in Eq. (4) has
nothing to do with $\theta^{}_{12}$ and $\delta^{}_{24}
-\delta^{}_{12}-\delta^{}_{14}$. This remarkable difference, which
cannot be eliminated even if one makes use of another
self-consistent parametrization of $V$, implies that both the X-ray
measurement and the neutrino capture experiment are important in
order to probe keV sterile neutrino DM and determine or
constrain its full parameter space.

\vspace{0.1cm}

\framebox{\bf 4} ~
To illustrate, we consider the tritium ($^3{\rm H}$) and ruthenium
($^{106}{\rm Ru}$) nuclei as the targets to capture keV sterile
neutrino DM. Their capture reactions have relatively large cross
sections, as one can see from Table 2 in the paper by Cocco {\it et
al} \cite{Cocco}. The typical values of $Q^{}_{\beta}$, $t^{}_{1/2}$
and $\sigma^{}_{\nu^{}_i} v^{}_{\nu^{}_i}$ for these two kinds of
nuclei are quoted as follows \cite{Cocco}: $Q^{}_{\beta} = 18.59
~{\rm keV}$, $t^{}_{1/2} = 3.8878\times 10^8 ~{\rm s}$ and
$\sigma^{}_{\nu^{}_i} v^{}_{\nu^{}_i}/c = 7.84\times 10^{-45} ~ {\rm
cm}^2$ for $^3{\rm H}$; or $Q^{}_{\beta} = 39.4 ~{\rm keV}$,
$t^{}_{1/2} = 3.2278\times 10^7 ~{\rm s}$ and $\sigma^{}_{\nu^{}_i}
v^{}_{\nu^{}_i}/c = 5.88\times 10^{-45} ~{\rm cm}^2$ for $^{106}{\rm
Ru}$, where $c$ is the speed of light. In addition, we adopt $|{\cal
M}|^2 \simeq 5.55$ for both $^3{\rm H}$ and $^{106}{\rm Ru}$
\cite{Weinheimer}. Our numerical analysis shows that the relative
values of $T^{}_e - Q^{}_\beta$ in the electron energy spectra of
the neutrino capture reaction and the $\beta$ decay are actually
insensitive to the inputs of $|{\cal M}|^2$, although ${\rm d}{\cal
N}^{}_\beta/{\rm d}T^{}_e$ itself is proportional to $|{\cal M}|^2$
and sensitive to its value.

We proceed to do a numerical calculation of ${\cal N}^{}_\nu$ and
${\rm d}{\cal N}^{}_\beta/{\rm d}T^{}_e$ for $^3{\rm H}$ and
$^{106}{\rm Ru}$ nuclei by using Eqs. (10) and (11). Because our
main concern is the signature of keV sterile neutrino DM, we
simply assume a normal mass ordering for three active neutrinos with
$m^{}_1 = 0$, $m^{}_2 = \sqrt{\Delta m^2_{21}} \simeq 8.7 \times
10^{-3}$ eV and $m^{}_3 = \sqrt{|\Delta m^2_{31}|} \simeq 4.9 \times
10^{-2}$ eV \cite{GG}
\footnote{A similar signature of $\nu^{}_4$ in the electron energy
spectrum can be obtained for the inverted or nearly degenerate mass
pattern of three active neutrinos, only if the input values of
those parameters associated with $\nu^{}_4$ are unchanged.}.
Moreover, we adopt the best-fit values $\theta^{}_{12} \simeq
34.4^\circ$, $\theta^{}_{13} \simeq 5.6^\circ$ and $\theta^{}_{23}
\simeq 42.9^\circ$ for the mixing angles among $\nu^{}_1$,
$\nu^{}_2$ and $\nu^{}_3$ \cite{GG} in the standard parametrization
of $V$. To illustrate the signature and background of $\nu^{}_4$, we
take two different mixing scenarios of sterile neutrinos as our
typical examples: (1) scenario A with $m^{}_4 = 2 ~{\rm keV}$ and
$|V^{}_{e4}|^2 \simeq 5 \times 10^{-7}$, compatible with the upper bound of
$s^2_{14} + s^2_{24} + s^2_{34}$ given in Eq. (5) so as to reveal the
most optimistic experimental prospect; (2) scenario B with
$m^{}_4 = 5 ~{\rm keV}$ and $|V^{}_{e4}|^2 \simeq 1
\times 10^{-9}$, consistent with the preliminary evidence for
keV sterile neutrino DM obtained recently in the Chandra X-ray Observatory \cite{Kusenko-DM}
\footnote{In this work we do not take into account the other preliminary
evidence for sterile neutrino DM with $m^{}_4 \sim 17.4$ keV and
$s^2_{14} + s^2_{24} + s^2_{34} \sim 10^{-12}$ \cite{Silk}, because it
implies $|V^{}_{e4}|^2 \sim 10^{-12}$ and leads to an extremely small
capture rate of $\nu^{}_4$ on $\beta$-decaying nuclei.}.

Our numerical results for the spectrum of the neutrino capture rate
against the $\beta$-decay background are shown in FIG. 2 and FIG. 3,
where a typical value of the finite energy resolution $\Delta$ ($=
2\sqrt{2\ln 2} \,\sigma \approx 2.35482 \,\sigma$) has been chosen
to distinguish the signal from the background. Furthermore, the
half-life $t^{}_{1/2}$ of target nuclei should be taken into
account because their number has been decreasing during the
experiment. We give a comparison between the result including the
finite half-life effect and that in the assumption of a constant
number of target nuclei for an experiment with the one-year exposure
time ($t=1\,{\rm year}$). To optimistically illustrate the signature
of keV sterile neutrino DM in this detection method, we assume
10 kg $^3{\rm H}$ and 1 ton $^{106}{\rm Ru}$ as the isotope
sources in our calculations.

FIG. 2 and FIG. 3 clearly show that the half-life effect is
important for the source of $^{106}{\rm Ru}$ nuclei but negligible
for the source of $^3{\rm H}$ nuclei. It may reduce about $30\%$ of
the neutrino capture rate on $^{106}{\rm Ru}$ in the vicinity of
$T^{}_e - Q^{}_\beta \simeq m^{}_4$. Hence this effect must be
included if the duration of such an experiment is comparable
with the half-life of the source. We see that smaller $m^{}_4$
requires a much better energy resolution (i.e., smaller $\Delta$).
The endpoint of the $\beta$-decay energy spectrum is sensitive to
$\Delta$, while the peak of the neutrino-capture energy spectrum is
always located at $T^{}_e \simeq Q^{}_{\beta} + m^{}_4$. So a comparison
between $\Delta$ and $m^{}_4$ can easily reveal the
signal-to-background ratio.
The required energy resolution to identify a signature of keV
sterile neutrino DM is of ${\cal O}(0.1) ~{\rm keV}$, which can
easily be reached in a realistic $\beta$-decay experiment (such as
the KATRIN experiment with $^3{\rm H}$ being the isotope source
\cite{Weinheimer}). Note that the endpoint location of the tritium
$\beta$-decay energy spectrum is slightly different from that of the
ruthenium $\beta$-decay energy spectrum for a given $\Delta$, simply
because they have different values of $Q^{}_\beta$. A large gap
between the location of the signature of $\nu^{}_4$ and the
$\beta$-decay endpoint in the electron recoil energy spectrum will
in practice make the signature itself almost independent of the
corresponding $\beta$-decay background.

The main problem which makes the observability of keV sterile
neutrino DM rather dim and remote is the tiny
active-sterile neutrino mixing angles. As one can see from FIG. 2
and FIG. 3, the capture rates of $\nu^{}_4$ on given $\beta$-decaying
nuclei are of ${\cal O}(1)$ in scenario A but
only of ${\cal O}(10^{-3})$ in scenario B. So a sizable
capture rate requires a great enhancement of other parameters in Eq.
(10), such as $N^{}_{\rm T}$, $n^{}_{\nu^{}_i}$ and
$\sigma^{}_{\nu^{}_i} v^{}_{\nu^{}_i}$. The largeness of a
combination of these factors may serve for a primary criterion for
us to search for the most promising target candidates. In Ref.
\cite{Cocco} some typical isotopes with sufficiently large values of
$\sigma^{}_{\nu^{}_i} v^{}_{\nu^{}_i} \, t_{1/2}$ have been listed
so as to pursue sufficiently large signal-to-noise ratios in the
detection of the C$\nu$B. Here our selection criterion for the
target nuclei is more or less the same. A re-analysis of all the
isotope sources shown in FIG. 5 of Ref. \cite{Cocco} is desirable in
order to optimize the detection of keV sterile neutrino DM.

Besides the $\beta$-decay background, two potential backgrounds
relevant to the capture of keV sterile neutrino DM on radioactive
$\beta$-decaying nuclei come from the capture of keV solar neutrinos
and from the coherent scattering of sterile neutrinos off the
electrons in the target nuclei (i.e.,
$\nu^{}_s + e^- \to \nu^{}_e + e^-$ or equivalently
$\nu^{}_4 + e^- \to \nu^{}_i + e^-$
for $i=1,2,3$). The standard solar model predicts the flux of solar
neutrinos around $E^{}_\nu \sim 10$ keV is about $10^{7} ~{\rm cm}^{-2}
\ {\rm s}^{-1}$ to $10^{8} ~{\rm cm}^{-2} \ {\rm s}^{-1}$ \cite{Bahcall},
and thus the number density of such keV solar neutrinos
(mainly $pp$ neutrinos) is around $10^{-4} ~{\rm cm}^{-3}$ to
$10^{-3} ~{\rm cm}^{-3}$, far below the inferred number density of keV
sterile neutrino DM $n^{}_{\nu^{}_4} \sim 10^{5} ~{\rm cm}^{-3}$
in our Galactic neighborhood. The keV solar neutrino background is
therefore negligible in the analysis of the capture of keV sterile
neutrino DM
\footnote{Even if this background were not negligibly small, it could be
rejected if the experiment is able to distinguish between solar and
anti-solar directions \cite{Ando}.}.
On the other hand, the $\nu^{}_4 + e^- \to \nu^{}_i + e^-$
scattering background is not expected to contaminate the signatures
of keV sterile neutrino DM in the capture reactions under discussion
either. As pointed out in Ref. \cite{Ando}, the momentum transfer to
the final-state $e^-$ and $\nu^{}_i$ is approximately equal to
$m^{}_4$. So the kinetic energy of
the outgoing electron is given by $m^2_4/(2 m^{}_e) \simeq 1 ~{\rm
eV} \ (m^{}_4/{\rm keV})^2$, which is of ${\cal O}(10)$ eV for
$m^{}_4$ to be a few keV. If the electron energy spectrum resulting
from this reaction is measured, it should sharply peak at $m^2_4/(2
m^{}_e)$ minus the atomic binding energy \cite{Ando}. In comparison,
the signature of keV sterile neutrino DM in a capture reaction will
peak at $T^{}_e \simeq Q^{}_\beta + m^{}_4 \gg m^2_4/(2 m^{}_e)$
as clearly shown in FIG. 2 and FIG. 3,
far away from the peak of the $\nu^{}_4$-$e^-$
scattering effect. Hence the latter is also negligible in the keV
sterile neutrino DM capture on $\beta$-decaying nuclei, although it
must exist for a neutrino detector. When $^{106}{\rm Ru}$ is used as
the target, however, one should take care of its metal property.
Only the electrons produced on the surface of the $^{106}{\rm Ru}$
target may assure the kinetic energy of the outgoing electron to be
equal or close to $T^{}_e \simeq Q^{}_\beta + m^{}_4$ in the presence
of a keV $\nu^{}_4$. An electron emitted from the interior of a ``thick"
$^{106}{\rm Ru}$ sample will have its energy altered by the
interaction with the ruthenium lattice. This effect can affect the
electron energy spectrum. Therefore, the actual line-like feature of
the $\nu^{}_4$ signature can only be expected from a ``thin" layer
of ruthenium, which is likely to make the size of an 1 ton $^{106}{\rm Ru}$ detector enormous
\footnote{We are grateful to the anonymous referee for calling our
attention to the metal property of $^{106}{\rm Ru}$ and its possible
effect on the capture of keV sterile neutrino DM.}.
How to build a realistic detector and how to do a feasible experiment
are still open questions.

\vspace{0.1cm}

\framebox{\bf 5} ~
In summary, we have argued that there might exist one or more keV
sterile neutrinos in the desert of the fermion mass spectrum. This
kind of warm DM can in principle be captured by means of radioactive
$\beta$-decaying nuclei, because the sterile component of $\nu^{}_e$
may leave a distinct imprint on the electron energy spectrum of the
capture reaction. For simplicity, only a single keV sterile neutrino
$\nu^{}_s$ or $\nu^{}_4$ is assumed in the present work. After
clarifying different active-sterile neutrino mixing factors for the
radiative decay of $\nu^{}_4$ and $\beta$ decays in a
self-consistent parametrization, we have carried out an analysis of
the signatures of $\nu^{}_4$ in the capture reactions $\nu^{}_e +
~^3{\rm H} \to ~^3{\rm He} + e^-$ and $\nu^{}_e + ~^{106}{\rm Ru}
\to ~^{106}{\rm Rh} + e^-$ against the $\beta$-decay backgrounds so
as to reveal a few salient features of this detection method. We
admit that it remains extremely difficult to detect keV sterile
neutrino DM even if it really exists, but we conclude that the
experimental approach under discussion should not be hopeless in the
long run.

\vspace{0.5cm}

We would like to thank W. Chao, W. Liao, W.L. Guo and S. Zhou for
very helpful discussions. One of us (Z.Z.X.) is also indebted to I.
Sogami for his warm hospitality at the Maskawa Institute for Science
and Culture in Kyoto, where this paper was written; and to J. Yokoyama 
for his warm hospitality during the COSMO/CosPA 2010 conference in
Tokyo, where this paper was finalized. He is also grateful to
K.K. Phua for his warm hospitality at the Institute of Advanced
Studies in Singapore, where this paper was revised. This work was 
supported in part by the National Natural Science Foundation of China 
under grant No. 10425522 and No. 10875131.


\begin{figure*}
\vspace{3.2cm}
\centering
\includegraphics[bb = 30 130 596 162,width=1.032 \textwidth]{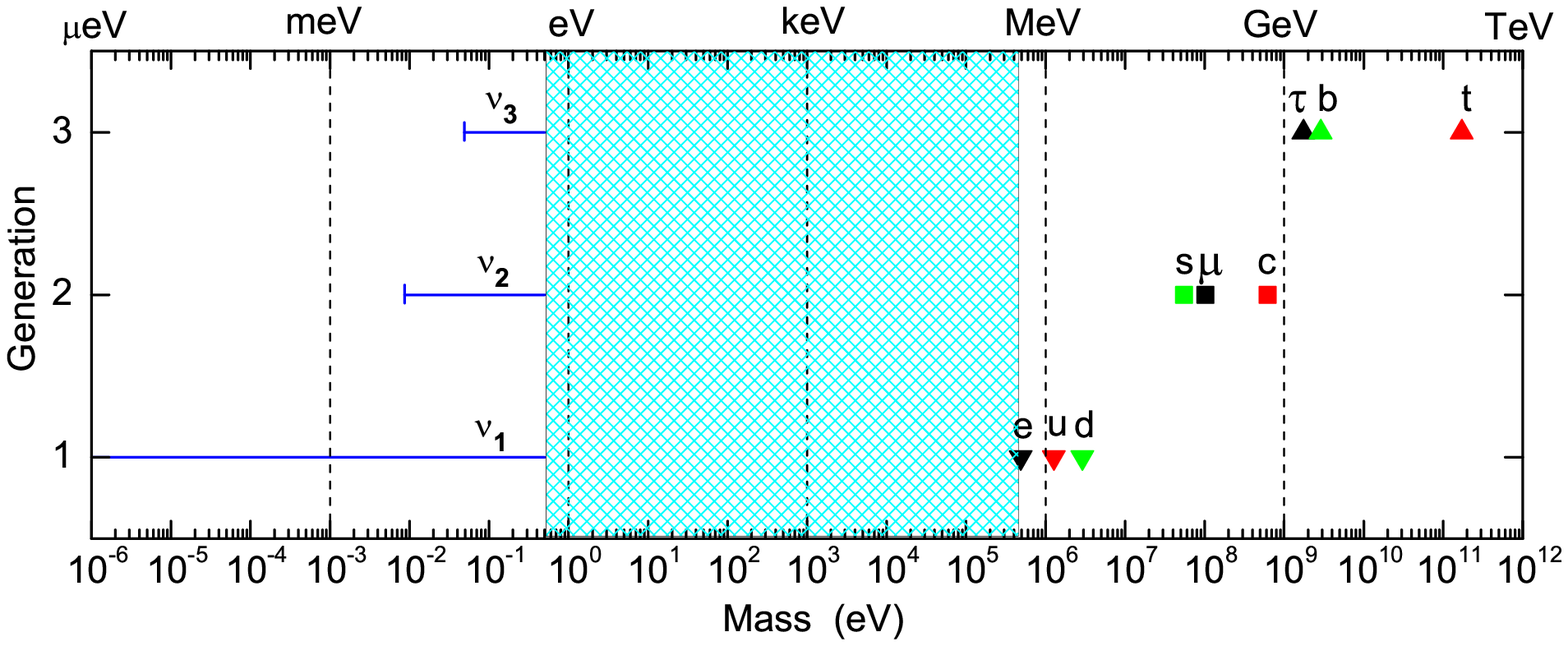}
\vspace{3.1cm} \caption{A schematic illustration of the flavor
``hierarchy" and ``desert" problems in the SM fermion mass spectrum
at the electroweak scale $M^{}_Z$, where the allowed ranges of
neutrino masses with a normal hierarchy are cited from Ref. [30],
and the central values of charged-lepton and quark masses are quoted
from Ref. [33].}
\end{figure*}

\begin{figure*}
\centering
\includegraphics[bb =140 120 480 680,width=0.7\textwidth]{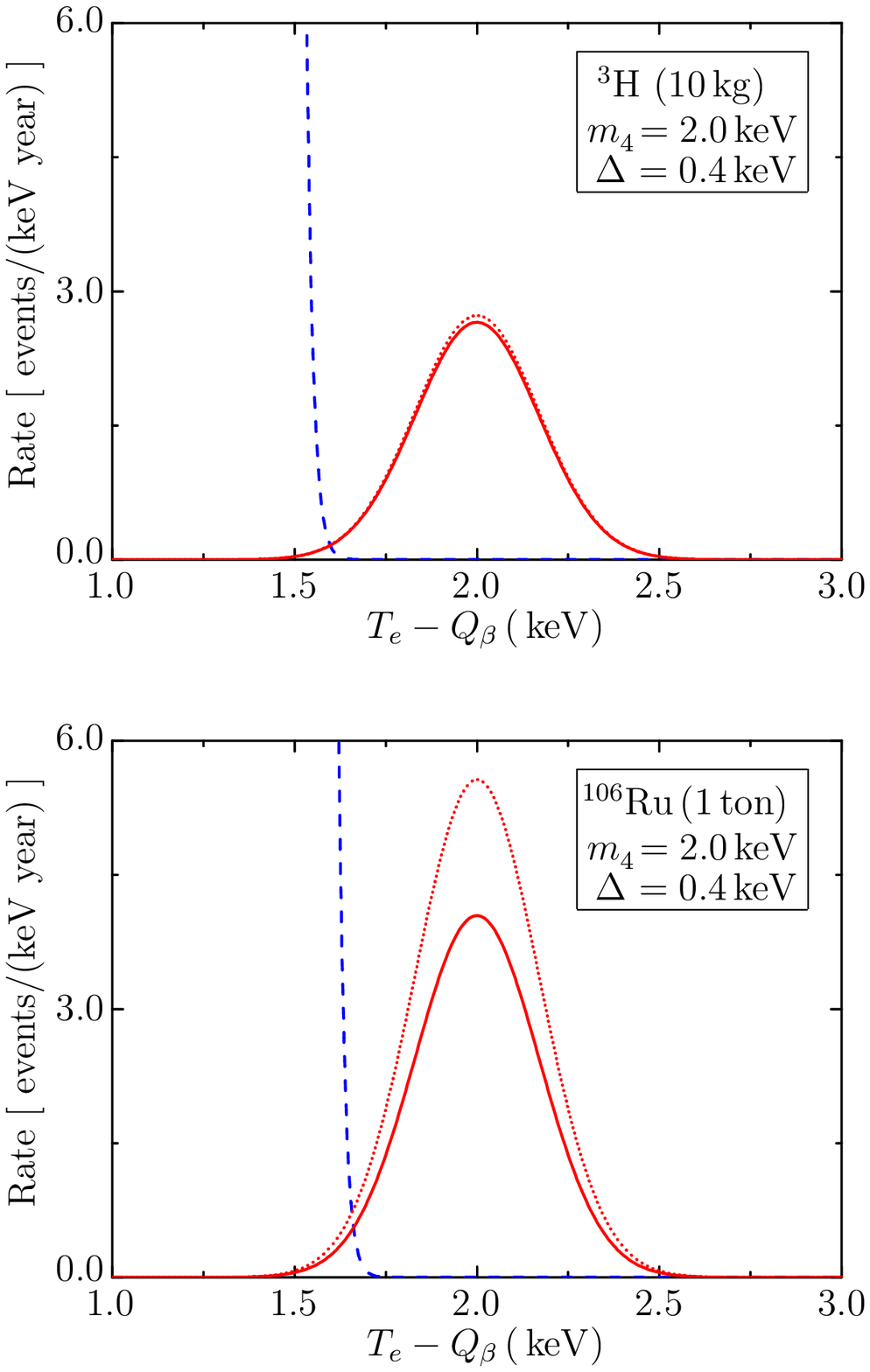}
\vspace{0.4cm} \caption{The capture rate of keV sterile
neutrino DM as a function of the kinetic energy $T^{}_e$ of
electrons, where $m^{}_4 = 2 ~{\rm keV}$ and $|V^{}_{e4}|^2 \simeq 5
\times 10^{-7}$ are typically input (scenario A). The solid (or dotted)
curves denote the signals with (or without) the half-life effect of
target nuclei. The dashed curve stands for the $\beta$-decay background.}
\end{figure*}

\begin{figure*}
\centering
\includegraphics[bb =140 120 480 680,width=0.7\textwidth]{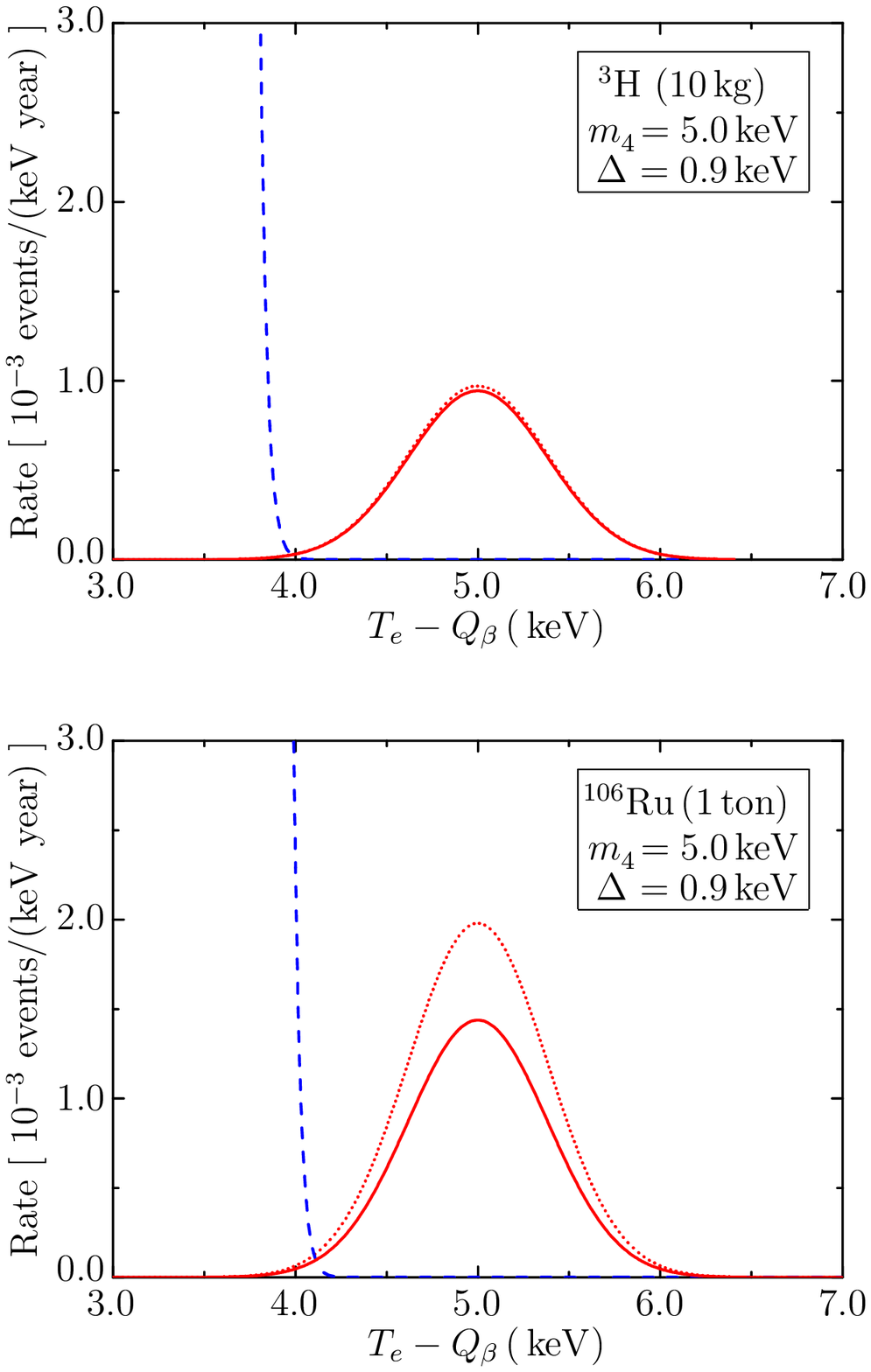}
\vspace{0.4cm} \caption{The capture rate of keV sterile
neutrino DM as a function of the kinetic energy $T^{}_e$ of
electrons, where $m^{}_4 = 5 ~{\rm keV}$ and $|V^{}_{e4}|^2 \simeq 1
\times 10^{-9}$ are typically input (scenario B). The solid (or dotted)
curves denote the signals with (or without) the half-life effect of
target nuclei. The dashed curve stands for the $\beta$-decay background.}
\end{figure*}

\end{document}